\documentclass[]{pasj01}
\draft

\Received{}
\Accepted{}
 
\begin{document} 

\title{ 
Hydrogen recombination near-infrared line mapping of Centaurus A with IRSF/SIRIUS }

\author{Risako \textsc{Katayama}\altaffilmark{1}%
}
\altaffiltext{1}{Graduate School of Science, Nagoya University, Furo-cho, Chikusa-ku, Nagoya, Aichi
464-8602, Japan}
\altaffiltext{2}{Institute of Liberal Arts and Sciences, Nagoya University, Furo-cho, Chikusa-ku, Nagoya, Aichi
464-8602, Japan}
\altaffiltext{3}{Institute of Space and Astronautical Science, Japan Aerospace Exploration Agency, Chuo-ku, Sagamihara, Kanagawa 252-5210, Japan}
\altaffiltext{4}{Institute of Liberal Arts and Sciences, Tokushima University, Minami Jousanjima-machi 1-1, Tokushima 770-8502, Japan}
\email{r.katayama@u.phys.nagoya-u.ac.jp}

\author{Hidehiro \textsc{Kaneda}\altaffilmark{1}}
\email{kaneda@u.phys.nagoya-u.ac.jp}

\author{Takuma \textsc{Kokusho}\altaffilmark{1}}


\author{Kumiko \textsc{Morihana}\altaffilmark{2}}
\author{Toyoaki \textsc{Suzuki}\altaffilmark{3}}
\author{Shinki \textsc{Oyabu}\altaffilmark{4}}
\author{Mitsuyoshi \textsc{Yamagishi}\altaffilmark{3}}
\author{Takuro \textsc{Tsuchikawa}\altaffilmark{1}}


\maketitle
\begin{abstract}
Centaurus A (Cen A) is one of the most famous galaxies hosting an active galactic nucleus (AGN), where the interaction between AGN activities and surrounding interstellar and intergalactic media has been investigated. Recent studies reported detections of the H$\alpha$ emission from clouds in the galactic halo toward the northeast and southwest of the nucleus of Cen A, suggesting that AGN jets may have triggered star formation there. We performed near-infrared line mapping of Cen A with the IRSF 1.4-m telescope, using the narrow-band filter tuned for Pa$\beta$, from which we find that the Pa$\beta$ emission is not detected significantly from either northeast or southwest regions. The upper limit of the Pa$\beta$/H$\alpha$ ratio in the northeast region is compatible with that expected for a typical H\,\emissiontype{II} region, in line with the scenario that AGNs have triggered star formation there. 
On the other hand, the upper limit of Pa$\beta$/H$\alpha$ in the southwest region is significantly lower than that expected for a typical H\,\emissiontype{II} region. A possibility to explain the low Pa$\beta$/H$\alpha$ ratio in the southwest region is the scattering of H$\alpha$ and Pa$\beta$ photons from the center of Cen A by dust grains in the halo clouds. 
From the upper limit of Pa$\beta$/H$\alpha$ in the southwest region, we obtain constraints on the dust size distribution, which is found to be compatible with those seen in the interstellar medium of our Galaxy.
\end{abstract}

\section{Introduction}
%
Active galactic nuclei (AGNs) are known to play an important role in the evolution of host galaxies through injecting kinetic energy into the surrounding interstellar and intergalactic media (ISM and IGM) via outflows or jets (\cite{HB}, and references therein); however it is yet to be understood how this AGN feedback mechanism affects the host galaxy. It has often been considered that AGN blows out cold gas from the host galaxy or prevents gas from cooling to suppress star formation (i.e., negative feedback; e.g., \cite{Fabian}; \cite{Harrison}). On the contrary, it is also suggested that the AGN activity can compress gas and trigger star formation (i.e., positive feedback; e.g., \cite{Silk}; \cite{Zinn}). Hence, the effect of AGN feedback is likely to be not straightforward (\cite{ZB}; \cite{Shin}).

The observational evidence of AGN-triggered star formation has been reported for nearby galaxies harboring AGNs, such as Minkowski's Object and 3C 285. Minkowski's Object is a star-forming peculiar object, located at the end of the radio jet from the nearby galaxy NGC 541, which has long been suggested that its star formation is triggered by the radio jet (\cite{Brodie}; \cite{vanB}; \cite{Salome}). At a distance of nearly $70\rm\ kpc$ from the galaxy center of 3C 285, \citet{vanBD} discovered a small H$\alpha$-emitting object near the eastern radio jet (3C 285/09.6), suggesting jet-induced star formation through compressing dense surrounding material.

Centaurus A (Cen A or NGC 5128), at a distance of 3.8 Mpc (\cite{Harris}), is the radio galaxy nearest to us, hosting an AGN which emits powerful radio and X-ray jets on the scales ranging from 1.35 kpc up to 250 kpc (\cite{Israel}). Thus, Cen A is an important target to understand interaction between AGN jets and the surrounding ISM and IGM. 
At a distance of approximately 8 kpc from the nucleus, optically bright filaments and possible young stars therein are observed in regions close to the northeastern radio jet, which could be the result of jet-ISM interaction (\cite{Blanco}; \cite{Osmer}; \cite{GP}; \cite{Morganti}; \cite{Rejkuba}; \cite{Crockett}). 
Additionally, recent studies report detection of hydrogen recombination line emissions in the galactic halo around the jets, indicating jet-induced star formation. In the northeast of Cen A, \citet{Santoro} find an H$\alpha$-emitting region at about 15 kpc away from the nucleus, suggesting that the AGN may have indeed triggered star formation there. On the opposite side of that region with respect to the nucleus, \citet{Keel} also find an H$\alpha$-emitting region in a halo cloud at about 12 kpc away from the nucleus along the axis of  the southwestern jet lobe, again suggesting a positive AGN feedback taking place in this region. 

On the other hand, there is a possibility of the scattering of H$\alpha$ photons from a host galaxy by dust grains in halo clouds. At about $4\rm\ kpc$ far from the galaxy disk of M 82, strongly polarized H$\alpha$ emission is observed, and its polarization pattern suggests that the H$\alpha$ emission originates from the dust scattering of H$\alpha$ photons coming from the central near-InfraRed (IR) nucleus (\cite{Scarrott}; \cite{Yoshida}). Therefore, in order to evaluate the effect of the AGN feedback accurately, it is important to investigate whether or not the hydrogen recombination line is truly of the origin of star formation therein.   

In the previous studies, only a small portion of Cen A was studied. In order to investigate the interaction between AGN feedback and the intergalactic clouds in Cen A, however, it is required to observe wider areas of Cen A along the axis of the jet with the hydrogen recombination lines tracing star-forming regions. Thus in the present study, we perform a wider area mapping of Cen A  along the axis of the jet using near-IR narrow-band filters. We obtain the physical parameters of the hydrogen recombination lines, combining our data with the data of the previous studies, to verify the possibility of the dust scattering in the halo clouds of Cen A. 
  
\section{Observations and data analysis}
Near-IR imaging toward Cen A was carried out in March, May and June 2019 with the SIRIUS (Simultaneous InfraRed Imager of Unbiased Survey; \cite{Nagashima}, \cite{Nagayama}) camera on the InfraRed Survey Facility (IRSF) 1.4-m telescope at  the South African Astronomical Observatory. We observed Cen A using the narrow-band filter tuned for 
Pa$\beta$ 1.282$\rm\ \mu m$ with the dithering of 10 frames and exposure time of 60 seconds per frame. The effective bandwidth of the filter for Pa$\beta$ is 0.029$\rm\ \mu m$. In addition, to subtract the Pa$\beta$ continuum, we also observed Cen A using the notch filter which has two peaks of transmittance at about 1.268 and 1.297$\rm\ \mu m$. The camera has a field of view of $\timeform{7'.7}\times \timeform{7'.7}$ with a pixel scale of $\timeform{0''.45}$. We observed five fields in total, and summarized the detail of the observation of each field in Table \ref{sum_ob}. For the continuum filter, we observed for twice as long total exposure times as those listed in Table \ref{sum_ob}.
\begin{table}[h]
  \tbl{Details of the IRSF observations for the present study.\label{sum_ob}}{%
  \begin{tabular}{cccccc}
      \hline
      Region & RA (J2000) & Dec (J2000) & Total exposure time (min) & Airmass & Date \\ \hline
      northeast & 13 26 20.00 & $-$42 50 00.0 & 30 & $1.02-1.24$ & 2019 May 17 \\
      northeast2 & 13 25 54.70 & $-$42 55 33.8 & 30 & $1.02-1.39$ & 2019 Mar 16 \\	
      center & 13 25 27.60 & $-$43 01 08.8 & 10 & $1.12-1.25$ & 2019 Mar 12 \\
      southwest2 & 13 25 01.70 & $-$43 04 21.1 & 30 & $1.24-1.98$ & 2019 Mar 17 \\
      southwest & 13 24 35.00 & $-$43 10 00.0 & 120 & $1.11-1.91$ &2019 Mar 15, Jun 15, 18, 20, 24 \\
	sky & 13 26 54.00 & $-$43 17 18.0 & 20 & $1.02-1.98$ & all days above  \\
      \hline
    \end{tabular}}\label{tab:first}
\begin{tabnote}
\end{tabnote}
\end{table}

The image data were processed with the standard data reduction by using the pipeline software of pyIRSF\footnote{https://sourceforge.net/projects/irsfsoftware/}, which includes dark subtraction, flat-fielding, sky subtraction and dithered-image-combining. We performed astrometric and photometric calibrations with the  2MASS Point Source Catalog (PSC; \cite{Skrutskie}), assuming that the magnitudes of point sources are the same between the $J$-band and the narrow-band images of Pa$\beta$ and the continuum. For photometry, we selected stars with the $J$-band fluxes higher than 12.5 mag which are not saturated, and with errors smaller than 0.05 mag from the 2MASS PSC.
Then, determining sky regions in each field and subtracting the sky emission from the images, we adjusted the sky level of each observed field to be zero.
Finally, we subtracted the Pa$\beta$-continuum image from the narrow-band image to obtain the Pa$\beta$ intensity map of Cen A.

\section{Results}
\begin{figure}
\begin{center}
\includegraphics[width=16cm]{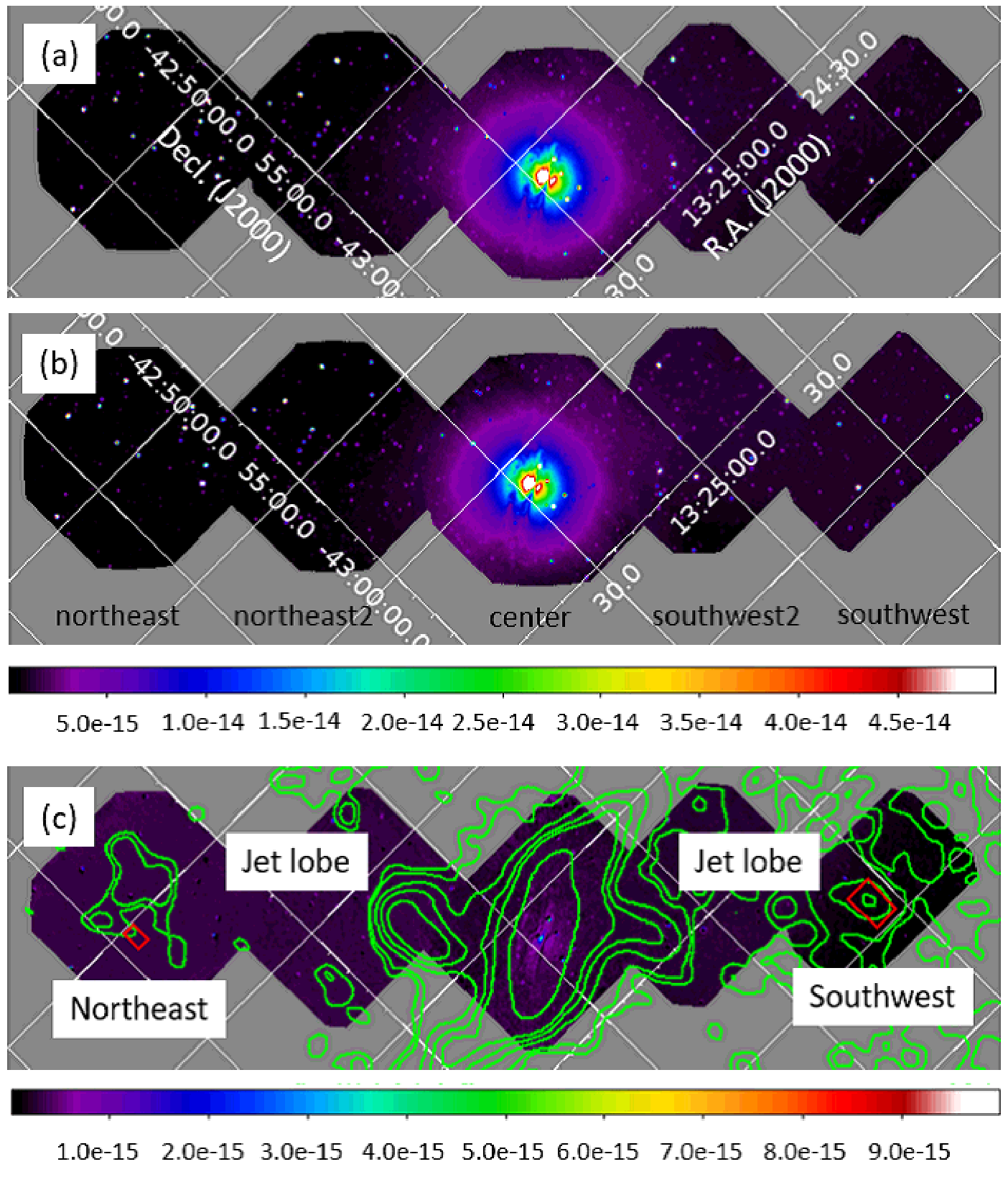} 
\end{center}
\caption{(a) Pa$\beta$+continuum, (b) continuum, and (c) continuum-subtracted Pa$\beta$ maps of Centaurus A. 
The maps have been smoothed with a Gaussian kernel of $\timeform{2''.7}$ in sigma, and color scales are given in units of $\rm erg\ s^{-1}\ cm^{-2}\ arcsec^{-2}$.
In panel (c), contours show the Herschel 500$\rm\ \mu m$ map from \citet{Auld}, where the contour levels correspond to $0.1, 6, 30,  60, 100, 500 \rm\ mJy\ beam^{-1}$. Red squares whose sizes are $30''\times45''$ and $60''\times90''$ indicate the northeast and the southwest regions where the H$\alpha$ emission is detected by \citet{Santoro} and \citet{Keel}, respectively. }\label{32map}
\end{figure}

\begin{figure}
\begin{center}
\includegraphics[width=16cm]{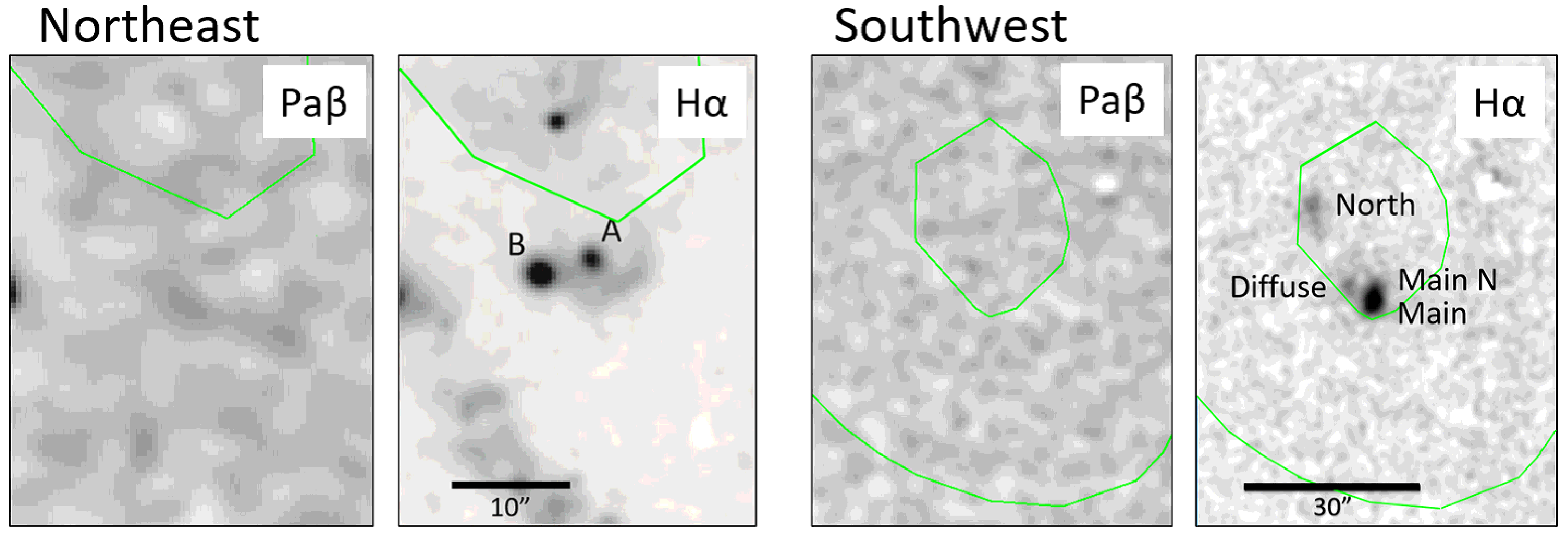} 
\end{center}
\caption{Comparison of the H$\alpha$ and the Pa$\beta$ line intensity maps of the northeast and the southwest regions of Centaurus A, which correspond to the red squares denoted in Figure \ref{32map}, overlaid with the same contours of the Herschel 500$\rm\ \mu m$ map as in Figure \ref{32map}. The Pa$\beta$ maps have been smoothed with a Gaussian kernel of $\timeform{1''.13}$ in sigma. The H$\alpha$ maps of the northeast and southwest regions are both taken from \citet{Keel}. $\copyright$ AAS. Reproduced with permission.}\label{NE-SW}
\end{figure}

Figure \ref{32map} shows the Pa$\beta$ intensity map of Cen A obtained with the narrow-band filters, from which we confirm the presence of the Pa$\beta$ emission extended along the galaxy structure, but not clearly associated with the jet lobes. Figure \ref{NE-SW} shows the close-up images of the northeast and southwest regions where the H$\alpha$ emission is detected in the previous studies. 
In the northeast region, \citet{Santoro} spectroscopically estimated the H$\alpha$ fluxes of Regions A and B to be $2.3\times10^{-15}$ and $3.8\times10^{-15}\rm\ erg\ s^{-1}\ cm^{-2}$, respectively. 
Similarly, in the southwest region, \citet{Keel} measured the H$\alpha$ flux densities from 4 H$\alpha$-bright local blobs of Main, Main $N$, Diffuse, and North with the integral-field spectroscopy, to be $9.3\times10^{-15}$,  $1.3\times10^{-15}$,  $1.1\times10^{-15}$ and $1.0\times10^{-15}\rm\ erg\ s^{-1}\ cm^{-2}$, respectively.
We performed aperture photometry with a radius of $\timeform{1.''5}$ for Regions A, B, and 4 H$\alpha$-bright local blobs using the present Pa$\beta$ map. As a result, we find that the Pa$\beta$ lines are not detected significantly from any of these regions (Table \ref{photometry flux}); the 3$\sigma$ upper limits of the Pa$\beta$ fluxes measured within the northeast and southwest regions are $7.2\times10^{-16}$ and $6.2\times10^{-16}\rm\ erg\ s^{-1}\ cm^{-2}$, respectively, where we use the sum of the photometry results for both regions.  

Assuming the Case B recombination with the electron density $n_e=10^2\rm\ cm^{-3}$ and electron temperature $T_e=10^4\rm\ K$ for a typical H\,\emissiontype{II} region, we estimate that the Pa$\beta$ flux from the H$\alpha$ flux of the previous study in the northeast region would be $3.4\times10^{-16}\rm\ erg\ s^{-1}\ cm^{-2}$. Thus the 3$\sigma$ upper limit of the observed flux ($<7.2\times10^{-16}\rm\ erg\ s^{-1}\ cm^{-2}$) is compatible with the Pa$\beta$ flux which is expected from the Case B recombination, indicating that the jet-induced star formation may have indeed taken place.
On the other hand, in the southwest region, we similarly estimate the Pa$\beta$ flux from the H$\alpha$ fluxes to be 7.2$\times10^{-16}\rm\ erg\ s^{-1}\ cm^{-2}$, assuming the Case B recombination, which is not compatible with the observed 3$\sigma$ upper limit in the southwest region ($<6.2\times10^{-16}\rm\ erg\ s^{-1}\ cm^{-2}$). Thus, the H$\alpha$ fluxes from this region may not be attributed to the emission of ionized gas in a star-forming region.

\begin{table}[h]
  \tbl{Results of the photometry of the Pa$\beta$ fluxes in the southwest and northeast regions of Cen A.\label{photometry flux}}{%
  \begin{tabular}{ccccc}
      \hline
	 $\bullet$ northeast component&  Region A & Region B && \\ \hline
	Observed Pa$\beta$ flux $\rm (erg\ s^{-1}\ cm^{-2})$&$(1.0\pm 1.7)\times 10^{-16}$& $(0.7\pm 1.7)\times 10^{-16}$&& \\ \hline \hline
	$\bullet$ southwest component & Main & Main $N$ & Diffuse & North  \\ \hline
	Observed Pa$\beta$ flux $\rm (erg\ s^{-1}\ cm^{-2})$& ($-0.6\pm 1.1)\times10^{-16}$ &  ($0.9 \pm 1.0)\times10^{-16}$ & ($8.0\pm 9.5)\times10^{-17}$ &  ($0.9\pm1.1)\times10^{-16}$\\ \hline
\end{tabular}}\label{tab:second}
\begin{tabnote}
\end{tabnote}
\end{table}

\newpage
\section{Discussion}
We obtain the observed upper limits of the Pa$\beta$/H$\alpha$ ratios of $1.2\times10^{-1}$ and $4.9\times10^{-2}$ in the northeast and southwest regions, respectively, which are compared with those predicted by models. In particular, we obtain the strongest constraint on the upper limit of the Pa$\beta$/H$\alpha$ ratio of 3.5$\times 10^{-2}$ for Main in the southwest region which is the brightest H$\alpha$ blob, and discuss its implications below.

First, we calculate the Pa$\beta$/H$\alpha$ ratios expected for the Case A and B recombinations for a typical H\,\emissiontype{II} region, as shown in Table \ref{CaseAB}, which are not consistent with the observed upper limit of the Pa$\beta$/H$\alpha$ ratio.
Even if we assume the Case B recombination for wide ranges of $n_e$ and $T_e$ of $10^2-10^6\rm\ cm^{-3}$ and $5000-20000\rm\ K$, respectively, the expected Pa$\beta$/H$\alpha$ ratios exceed $5.3\times10^{-2}$ (\cite{OF}).
Thus, we find that 
neither Case A nor B satisfies the observed upper limit of the Pa$\beta$/H$\alpha$ ratio in the southwest region. 
 Then we consider the effect of dust extinction; however it would further increase the Pa$\beta$/H$\alpha$ ratio to be observed, because the Pa$\beta$ emission is less attenuated by dust compared to the H$\alpha$ emission. 
%

It is possible that the jet-driven shock excites the optical emission line. For example, \citet{Sutherland} suggest that the inner emission-line filaments in Cen A at approximately 8 kpc away from the nucleus of Cen A can be produced by the interaction between the jet and the surrounding ISM. 
Actually we estimate the Pa$\beta$/H$\alpha$ ratio assuming shock excitation with $n_e$ and $T_e$ of $10^2-10^4$ and $500\rm\ K$, respectively, to find that the resultant Pa$\beta$/H$\alpha$ ratio is larger than $2.1\times 10^{-2}$ (\cite{OF}), which is compatible with the observed 3$\sigma$ upper limit of the Pa$\beta$/H$\alpha$ ratio.
In this shock excitation model, however, a strong radio jet needs to be present in the vicinity of the emission-line filaments. In the northeast region of Cen A, \citet{Santoro15} claim that the jet-cloud interaction is unlikely to be important because the jet is too diffuse in this region. Thus, the jet-driven shock excitation may not be taking place in the southwest region as well, where the galactic cloud is 12$\rm\ kpc$ away from the nucleus and the jet is as diffuse as in the northeast region.

Yet another possibility to explain the observed upper limit of the Pa$\beta$/H$\alpha$ ratio in the southwest region is the dust scattering of the H$\alpha$ and Pa$\beta$ photons emitted from the center of the host galaxy (Table \ref{CaseAB}). 
In this case, the scattering efficiency of Pa$\beta$ photons is expected to be significantly lower than that of H$\alpha$ photons. 
Indeed, in the southwest region, \citet{Auld} identified a dust cloud in the 500-$\rm \mu m$ emission, who excluded the possibility that the 500-$\rm \mu m$ emission is due to synchrotron radiation.
A scattering scenario would also be useful to explain the low dust temperature and the lack of UV emission observed by \citet{Auld} for the dust cloud in the southwest region. On the other hand, \citet{Keel} find that the optical spectrum in the southwest region is consistent with photoionization in typical H\,\emissiontype{II} regions. Since scattering preserves local line ratios, it would suggest scattering not so much of AGN radiation as that from the inner starburst disk.

The dust-scattered H$\alpha$ emission from the host galaxy is actually observed in the galactic halo of M 82 (\cite{Yoshida}). 
Assuming that it is also the case in Cen A, we measured the Pa$\beta$ flux of the host galaxy with an aperture radius of $\timeform{2'.5}$. 
The resultant Pa$\beta$ flux is $(8.0\pm 0.7)\times 10^{-13}\rm\ erg\ s^{-1}\ cm^{-2}$ received by us and $(8.0\pm 0.7)\times 10^{-8}\rm\ erg\ s^{-1}\ cm^{-2}$ by dust in the southwestern cloud with the distance of 12 kpc from the galaxy to the cloud. Using these results, we calculate the H$\alpha$ flux of the host galaxy for the Case B recombination, and then we estimate the Pa$\beta$ and H$\alpha$ fluxes scattered by dust in the southwestern cloud.

First, in order to get a rough picture, we tentatively use silicate and graphite dust of a constant size of $0.35\rm\ \mu m$ in radius, for which the scattering coefficients at wavelengths of the Pa$\beta$ and H$\alpha$ lines, $Q_{\rm Pa\beta}$ and $Q_{\rm H\alpha}$, are taken from \citet{LD}. As a result, we find that the estimated Pa$\beta$/H$\alpha$ ratio, both scattered by the silicate and graphite dust of $0.35\rm\ \mu m$ in size in the southwestern cloud, would be $(3.0\pm 0.3)\times10^{-2}$ and $(5.2\pm 0.4)\times10^{-2}$, respectively (Table \ref{CaseAB}). Hence the silicate dust scattering is compatible with the 3$\sigma$ upper limit of the observed Pa$\beta$/H$\alpha$ ratio in the southwest region, whereas the graphite dust scattering is not.

\begin{table}[h]
  \tbl{Comparison of the observed and model-predicted Pa$\beta$/H$\alpha$ ratios.\label{CaseAB}}{%
  \begin{tabular}{cccccc}
      \hline
     	 & Observed $3\sigma$ upper limit &Case A\footnotemark[$*$]& Case B\footnotemark[$\dag$]  & \multicolumn{2}{c}{dust scattering\footnotemark[$\ddag$]} \\
	&&&&silicate&graphite \\ \hline
			Pa$\beta$/H$\alpha$&$< 3.5\times10^{-2}$ (Main southwest)&$7.6\times10^{-2}$&$5.7\times10^{-2}$&$(3.0\pm 0.3)\times10^{-2}$&$(5.2\pm 0.4)\times10^{-2}$ \\
	\hline
    \end{tabular}}\label{tab:third}
\begin{tabnote}
\footnotemark[$*$] low density limit, $T_e=10^4\rm\ K$(\cite{OF})\\ 
\footnotemark[$\dag$] $n_e=10^2\rm\ cm^{-3}, T_e=10^4\rm\ K$(\cite{OF})\\ 
\footnotemark[$\ddag$] Scattering efficiency from \citet{LD}. A single dust size of $0.35\rm\ \mu m$ is assumed. The scattered Pa$\beta$/H$\alpha$ photons originate from the Case B recombination lines from the host galaxy.\\
\end{tabnote}
\end{table}

Then we assume a more realistic dust size distribution such as that in \citet{MRN} with $Q_{\rm Pa\beta}$ and $Q_{\rm H\alpha}$ in  \citet{LD}, the result of which is shown in Table \ref{MRN scattering}. We adopt the ranges of silicate and graphite dust sizes of $0.025-0.25\rm\ \mu m$ and $0.005-1\rm\ \mu m$, respectively (MRN77). 
Table \ref{MRN scattering} also shows the results for power-law indices different from that of MRN77. As compared to Table \ref{CaseAB}, the Pa$\beta$/H$\alpha$ ratios in Table \ref{MRN scattering} are relatively small due to the presence of dust smaller than 0.35 $\rm \mu m$ in size, and therefore the observed result favors the presence of such small-size dust.
\begin{table}[h]
  \tbl{Pa$\beta$/H$\alpha$ ratios expected from dust scattering with different power-law indices of the dust size distribution.\label{MRN scattering}}{%
  \begin{tabular}{ccccccc}
      \hline
	power-law index & Observed $3\sigma$ upper limit & $-$3.5 (MRN77) &$-$2.5& $-$4.0 \\ \hline
	Pa$\beta$/H$\alpha$&$< 3.5\times10^{-2}$ (Main southwest)&$1.1\times10^{-2}$&$1.8\times10^{-2}$&$8.7\times10^{-3}$ \\ \hline
    \end{tabular}}\label{tab:forth}
\begin{tabnote}
\end{tabnote}
\end{table}
Consequently, the size distribution following the MRN77 power-law function can reproduce the Pa$\beta$/H$\alpha$ ratio compatible with the observed upper limit.
As shown in Table \ref{MRN scattering}, as the power-law index is steeper, the compatibility with our observational result is more secure.
Actually, in the galactic halo, \citet{HL} show a relative high abundance in small grains with a radius of less than 0.03$\rm\ \mu m$ with their numerical simulation, which may be caused by dust shattering during the galactic outflow processes. 
In order to give a stronger constraint on the dust size distribution in this region, we need further observations of Pa$\beta$ with higher sensitivity.

\section{Conclusion}
We have conducted near-IR line mapping of Cen A with IRSF, using the narrow-band filter tuned for the Pa$\beta$ line to study the interaction between the AGN activity and the surrounding IGM in the northeast and the southwest regions of the halo of Cen A. In the previous studies, the H$\alpha$ emission is detected in these regions (\cite{Santoro}; \cite{Keel}), suggesting that AGN-triggered star formation may have taken place therein. 
Our observations, however, do not detect the Pa$\beta$ emission, and in the southwest region, the upper limit of the Pa$\beta$/H$\alpha$ ratio is significantly lower than that expected for a typical H\,\emissiontype{II} region. 
One possibility to explain the observed ratio is the scattering of H$\alpha$ and Pa$\beta$ photons from the center of the host galaxy by dust in the galactic halo. If this is the case, the H$\alpha$ fluxes detected in these regions are not relevant to the AGN-induced star formation, but just reflecting the presence of dusty clouds where the H$\alpha$ photons originating from the galactic center of Cen A are scattered by the dust grains therein. 
According to the scattering scenario, the observed 3$\sigma$ upper limit of the Pa$\beta$/H$\alpha$ ratio is consistent with the size distribution of the interstellar dust in our Galaxy.
Finally our result suggests the importance of observing Pa$\beta$ emission in addition to H$\alpha$ to identify the origin of the H$\alpha$ emission in the halo of a galaxy.
\begin{ack}
We are grateful to the referee for giving us useful comments. The IRSF project is a collaboration between Nagoya University and the SAAO supported by the Grants-in-Aid for Scientific Research on Priority Areas (A) (Nos. 10147207 and 10147214) and Optical \& Near-Infrared Astronomy Inter- University Cooperation Program, from the Ministry of Education, Culture, Sports, Science and Technology (MEXT) of Japan and the National Research Foundation (NRF) of South Africa. K.M. is financially supported by MEXT/JSPS KAKENHI Grand number 17K18019.
\end{ack}

%
%
%
%


\end{document}